\documentclass[aps,prb,superscriptaddress,twocolumn,floatfix]{revtex4-2}
\usepackage[utf8]{inputenc}
\newcommand{\be}{\begin{equation}}
\newcommand{\ee}{\end{equation}}
\newcommand{\bea}{\begin{eqnarray}}
\newcommand{\eea}{\end{eqnarray}}
\usepackage{amsmath}
\usepackage{amssymb}
\usepackage{amstext}
\usepackage{mathrsfs}
\usepackage{graphicx}
\usepackage{hyperref}
\usepackage{makecell}
\usepackage{simpler-wick}
\usepackage{float}
\usepackage{xcolor}
\usepackage{ulem}
\usepackage{comment}
\hypersetup{
  colorlinks   = true, %Colours links instead of ugly boxes
  urlcolor     = blue, %Colour for external hyperlinks
  linkcolor    = blue, %Colour of internal links
  citecolor   = red %Colour of citations
}

\begin{document}

\title{Optical conductivity of the two dimensional Hubbard model: vertex corrections, emergent Galilean invariance and the accuracy of the single-site dynamical mean field approximation. }
\author{Anqi Mu}
\affiliation{Department of Physics, Columbia University, New York, NY 10027, USA}
\author{Zhiyuan Sun}
\affiliation{Department of Physics, Harvard University, Cambridge, MA 02138, USA}
\author{Andrew J. Millis}
\affiliation{Department of Physics, Columbia University, New York, NY 10027, USA}
\affiliation{Center for Computational Quantum Physics, Flatiron Institute, 162 5th Avenue, New York, NY 10010, USA}

\date{\today}

\begin{abstract}
We compute the frequency dependent conductivity of the two dimensional square lattice Hubbard model at zero temperature as a function of density to second order in the interaction strength, and compare the results to the predictions of single-site dynamical mean field theory computed at the same order.  We find that despite the neglect of vertex corrections, the single site dynamical mean field approximation produces semiquantitatively accurate results for most carrier concentrations, but fails qualitatively for the nearly empty or nearly filled band cases where the model exhibits an emergent Galilean invariance. The DMFT approximation  also becomes qualitatively inaccurate very near half filling if nesting is important. %The effective Galilean invariant theory emerging at low density is however characterized by a current channel Landau parameter $F_{1S}/2\neq \frac{m^*}{m}-1$ and a correction to the total spectral spectral weight proportional to the square of the density.
\end{abstract}

\maketitle
\section{Introduction}
\label{sec:intro}
The single-site dynamical mean field theory (DMFT) \cite{georges1996dynamical} is a widely used approximate method for computing properties of strongly correlated electron systems. It becomes exact for lattice models in an infinite coordination number (infinite dimensional) limit,  is believed to provide a reasonable approximation to many aspects of the physics of  interacting systems on finite dimensional lattices, and can be combined with band theory calculations to provide a chemically realistic description of the properties of wide classes of quantum materials \cite{Kotliar2006a,amadon2008plane}. The single site DMFT approximation treats the electron self energy as local in an appropriate orbital basis, and the magnitude of the errors induced by this approximation is the subject of active investigation. Much has been understood about regimes of temperature and carrier concentration where momentum-dependent correlations are important for static equilibrium  properties \cite{Gull10,LeBlanc15,Wietek21} but in the context of transport properties, while interesting results have been obtained in a high temperature limit \cite{Peripelitsky16,Vranic20}, the situation is less well understood.

A transport property of particular interest  is the optical conductivity $\sigma(\omega)$, which is the linear response function relating an applied long wavelength transverse electric field $E(\omega)$ to a measured current $j(\omega)$. $\sigma(\omega)$ is of fundamental interest because it reveals how electronic motion is affected by the combination of electron-electron interactions and ionic potential and of practical interest because it is a convenient and widely studied experimental probe of quantum materials.

In the standard diagrammatic analysis, theoretical computation of the conductivity requires knowledge of both the self energy $\Sigma(p,\omega)$, expressing how carriers move and are scattered under the influence of interactions, and the ``vertex correction" expressing interaction contributions to the coupling to external fields and also encoding the physics of conservation laws.  In a Galilean invariant system (in other words, a system with a continuous translation invariance and a $p^2/2m$ electron dispersion), the vertex correction exactly cancels the self energy\cite{Pines66,Negele98}, so that the conductivity takes exactly the free-particle form, independent of interactions. The conductivity vertex correction is related to the momentum dependence of the electron self energy  and vanishes in the single-site dynamical mean field approximation \cite{georges1996dynamical}.  In any reasonable model, as the band filling tends zero, the dispersion tends to $\varepsilon_p\sim p^2$, the  non-interacting physics becomes approximately Galilean invariant and one may expect that vertex corrections, neglected in the dynamical mean field theory, become particularly  important to the computation of the conductivity in the low density limit. 

At frequencies well below the lowest interband transition the conductivity may be written as $\sigma\propto 1/(-i\omega(1+\Lambda(\omega,T))+\Gamma(\omega,T))$ where $\Lambda$ and $\Gamma$ represent interaction-induced mass renormalization and scattering and vanish as the interaction tends to zero. The presence of $\Lambda,\Gamma$ in the denominator means that the conductivity is in general not perturbatively accessible; however at $T=0$ and at weak correlations $\Lambda\ll1$ and $\Gamma\ll \omega$, implying that a perturbative treatment is possible. In this paper we exploit this fact to compute  the frequency dependent conductivity of the two dimensional square lattice Hubbard model at zero temperature as a function of carrier density perturbatively to second order in the interaction, obtaining results which are exact to order $U^2$ with corrections of higher order. We compare these results  to the predictions of DMFT to the same order in $U$. A similar method was applied to Dirac and related systems by Sharma, Principi and Maslov \cite{sharma2021optical}. 

The rest of the paper is organized as follows. In Section \ref{sec:form} we introduce the formalism and the model. Section \ref{sec:K} presents results for the total spectral weight (integral of the real part of the conductivity over all frequencies). Section \ref{sec:ffp} presents results for the functional form of $\sigma(\omega)$ at nonzero frequency. Section \ref{sec:dw} gives the Drude weight correction. Section \ref{sec:conclusion} is a conclusion.

\section{Formalism}
\label{sec:form}
\subsection{Conductivity: definitions}
\label{subsec:conductivity}
We consider a system described by a Hamiltonian $\hat{H}[\mathbf{A}]$ depending on a spatially uniform time-dependent vector potential $\mathbf{A}$ related to the electric field as $\mathbf{E}=-\partial_t\mathbf{A}$. (In this paper we choose units such that the speed of light, Planck constant $\hbar$ and the electric charge are set to unity). The current operator $\mathbf{j}=-\frac{\delta \hat{H}}{\delta \mathbf{A}}$, the minimal coupling relation $k\rightarrow k-A$ and standard linear response arguments \cite{Pines66,jaklivc2000finite} yield the elements of the temperature $T=0$ conductivity tensor $\sigma^{\alpha,\beta}$ relating the current in the $\alpha$ direction induced by a field in the $\beta$ direction) as
\bea
\sigma^{\alpha\beta}(\omega)=\frac{1}{i\omega}\left[ -K^{\alpha\beta}+\chi_{jj}^{\alpha\beta}(\omega)\right], 
\label{eq:sigma}
\eea
where
\begin{equation}
    K^{\alpha\beta}=\Big \langle\frac{\delta ^2 \hat{H}}{\delta A^{\alpha}\delta A^{\beta}}\Big\rangle,
    \label{eq:K}
    \end{equation}
    and the current-current correlation function is
\begin{equation}
\chi_{jj}^{\alpha\beta}(t-t')=i \left\langle\left[\hat{j}^{\alpha}(t),\hat{j}^{\beta}(t')\right] \right\rangle\theta(t-t').
\label{eq:chidef}
\end{equation}
The expectation values are taken in the   ground state of the model. We will specialize to high symmetry situations (in the two dimensional case, this would include the O(2) symmetry of free electrons,  $C_4$ symmetry of electrons on a square lattice, and the $C_6$ symmetry of the hexagonal lattice) where the conductivity is proportional to the unit tensor and for explicit calculations take the electric field and current operator to be in the x direction.

The real $\sigma_1$ and imaginary $\sigma_2$ parts of the conductivity obey a Kramers-Kronig relation,
\begin{equation}
    \sigma_2(\omega)=\mathcal{P}\int\frac{dx}{\pi}\frac{\sigma_1(x)}{\omega-x}.
    \label{eq:KKsigma}
\end{equation}
Because the current-current correlation function vanishes rapidly as $\omega\rightarrow\infty$, we obtain by comparing the $\omega\rightarrow\infty$ limits of Eq.~(\ref{eq:sigma}) and Eq.~(\ref{eq:KKsigma}),
\begin{equation}
    K=\int\frac{dx}{\pi}\sigma_1(x).  
    \label{eq:KK}
\end{equation}

In a system with a discrete translational invariance at $T=0$, it may be that 
\begin{equation}
    \lim_{\omega\rightarrow 0}\left(K-\chi_{jj}(\omega)\right)\equiv K_D\neq 0,
    \label{eq:Adef}
\end{equation}
so that the low frequency limit of the conductivity may be written as
\begin{equation}
   \sigma(\omega)=\frac{ -K_D}{i\omega}+\sigma_{reg}(\omega),
   \label{eq:sigmalow}
\end{equation}
defining the ``Drude weight" $K_D$. Here $\lim_{\omega\rightarrow 0}\omega\sigma_{reg}(\omega)=0$. The term proportional to $K_D$ represents the fraction of the carriers that may be freely accelerated by an electric field.

In a Galilean-invariant system the current operator is identical to the momentum operator and commutes with the Hamiltonian, so $\chi_{jj}=0$ independent of interactions and $K=K_D=n/m$ (we have set the charge equal to unity).  In a continuum but not Galilean invariant model (e.g. the band theory problem of electrons in the presence of a periodic array of ions) K defined as the integral of the conductivity over all frequencies (i.e. including all interband transitions and transitions to the continuum) is equal to n/m independent of interactions but $K_D<K$, reflecting the fact that the ionic potential prevents some fraction of the electrons from freely accelerating in an applied dc field. Our interest in this paper is in tight binding models, which describe only a subset of the orbitals in a real solid. The conductivity in this case refers only to those transitions involving states described by the tight binding model and  $K$ will depend on interactions  as well as on band filling. 

In summary, we may characterize the conductivity by three quantities: the ``total spectral weight" $K$ (Eq.~(\ref{eq:KK})), the ``Drude weight'' of freely accelerating carriers $K_D$ (Eq.~(\ref{eq:Adef})) and the form of the frequency dependent conductivity $\chi_{jj}/\omega$.  In the rest of this paper we investigate, within a perturbative approximation to a simple model, the extent to which the dynamical mean field approximation accurately captures the magnitude and frequency dependence of these effects.

\subsection{Calculational formalism}

As mentioned in the introduction, at frequencies sufficiently less than the lowest interband transition energy the conductivity may alternatively be written (neglecting the interband contribution to the low frequency dielectric constant) as 
\bea
\sigma(\omega) = \frac{K}{-i\omega(1+\Lambda(\omega))+\Gamma(\omega)}, 
\label{eq:sigmaasmemoryfunction}
\eea
where  the  ``memory function"  $-i\omega\Lambda(\omega)+\Gamma(\omega)\equiv K \sigma(\omega)^{-1}+i\omega$. Because the inverse conductivity is a physically well defined response function, $\omega\Lambda$ and $\Gamma$ also obey a Kramers-Kronig relation, and following from the properties of $\sigma$ we see that both $\Lambda$ and $\Gamma$ are even functions of $\omega$.  %\ajm{\bf already said in intro}The real part of $M(\omega)$ can be thought of as a frequency dependent scattering rate and the imaginary part as $\omega$ times a frequency dependent mass. 
At $T=0$ in a Fermi liquid  with discrete translational invariance we expect $\displaystyle{\lim\limits_{\omega\to 0}\Lambda}= A+\mathcal{O}\omega^2$ and $\displaystyle{\lim\limits_{\omega\to 0}\Gamma}=B\omega^2+\mathcal{O}\omega^4$. Both  $\Lambda$ and $\Gamma$ vanish at high frequencies and also vanish in the non-interacting limit, so for small interactions 

\bea
\sigma(\omega) = \frac{-K}{i\omega(1+\Lambda(\omega))}+\frac{K\Gamma(\omega)}{\omega^2},
\label{eq:sigmaasmemoryfunctionexpand}
\eea
implying
\begin{eqnarray}
K_D&=&\frac{K}{1+A}\approx K(1-A),
\label{KDPT} 
\\
Re\left[\sigma_{reg}(\omega)\right]&=&\frac{K\Gamma(\omega)}{\omega^2}, 
\label{sigmaregpt}
\end{eqnarray}
and, by comparing to Eq.~(\ref{eq:sigma}),
\bea
\Gamma(\omega)=\frac{\omega Im[\chi_{jj}(\omega)]}{K}.
\label{eq:Mandchijj}
\eea
Thus a perturbative calculation of $\chi_{jj}$ provides a perturbative calculation of $\Gamma$ which in the weak coupling, $T=0$  limit  provides a complete description of the dissipative part of the conductivity. At nonzero  temperature, this method  would fail at low frequencies because $\Gamma$ would have a term proportional to $T^2$ so $\Gamma/\omega$ would not be small below a small frequency of the order of the square of the interaction times the  temperature. Some kind of resummation would have to be performed, but this is not considered here. 

We calculate the current-current correlation function using the force-force method \cite{mahan}. The essential idea is to integrate by parts, noting that time derivatives correspond to commutators with the Hamiltonian and that the non-interacting term commutes with the current. 

The result is compactly expressed in terms of the force operator $\hat{F}^\alpha=\left[\hat{H},\hat{j}^\alpha\right]$ as (see Appendix~\ref{app:pert})
\begin{widetext}
\bea
    \chi_{jj}^{xx}(i\omega_n)=-\frac{1}{(i\omega_n)^2}\left\langle \left[\hat{j}^x(0),\frac{d\hat{j}^x(\tau)}{d\tau}\bigg|_{\tau=0}\right]\right\rangle -\frac{1}{(i\omega_n)^2}\int_{0}^{\beta}d\tau e^{i\omega_n\tau}\left\langle T_{\tau}\hat{F}^x(\tau)\hat{F}^x(0)\right\rangle.
\label{eq:ordering1}
\eea
\end{widetext}
The first term is real and does not contribute to the absorptive part of the conductivity; we will focus on the second term. For notational convenience we have written the formulas on the imaginary time axis.

In the dynamical mean field formalism,  vertex corrections vanish and the current-current correlation function is given \cite{georges1996dynamical} as a convolution of two electron Green functions and two velocity operators,
\bea
\chi_{jj}^{xx}(i\omega_n)&=& -T\sum_{\nu_n,\sigma}\sum_k v_kG_{\sigma}(k,i\omega_n+i\nu_n)v_kG_{\sigma}(k,i\nu_n). \nonumber \\
\eea

\subsection{Hubbard model}
\label{subsec:hubbard}
We specifically study the one-band two dimensional square lattice Hubbard model with nearest neighbour hopping. The Hamiltonian $\hat{H}$ contains a quadratic (hopping) part $\hat{T}$ and an interaction part $\hat{H_I}$, where
\bea
\hat{T} &=& -\sum_{ij\sigma}t_{ij}c_{i\sigma}^{\dagger}c_{j\sigma}=\sum_{k\sigma}E_kc_{k\sigma}^{\dagger}c_{k\sigma}, \nonumber \\ 
\hat{H_I}&=& \sum_{i}Un_{i\uparrow}n_{i\downarrow}=U\sum_{kk'q}c_{k\uparrow}^{\dagger} c_{k+q\uparrow}c_{k'\downarrow}^{\dagger}c_{k'-q\downarrow}.
\label{eq:HHubard}
\eea
For the nearest neighbour hopping $E_k=-2t\left(\cos(k_xa)+\cos(k_ya)\right)$.  We set $t=1$ and lattice constant $a=1$ throughout the paper. The carrier concentration ranges from $0$ to $2$. 

The total spectral weight is found by substituting the first line of Eq.~(\ref{eq:HHubard}) into Eq.~(\ref{eq:K}) and using the momentum-space form of the dispersion given below Eq. ~(\ref{eq:HHubard}) and the minimal coupling $k\rightarrow k-A$:
\begin{equation}
    K=\sum_{k,\sigma}2t\cos(k_x)\left<c^\dagger_{k\sigma}c_{k\sigma}\right>. 
    \label{eq:KHubbard}
\end{equation}

The force operator is, explicitly
\bea
\hat{F}^x &=&[\hat{H},\hat{j}^x] \nonumber \\
&=& U\sum_{kk'q}c_{k\uparrow}^{\dagger}c_{k'\downarrow}^{\dagger}c_{k'-q\downarrow}c_{k+q\uparrow}(v_{k+q}+v_{k'-q}-v_{k}-v_{k'}),\nonumber \\
\label{eq:FHubbard}
\eea
where
\begin{equation}
    v_k=\frac{\partial E_k}{\partial k_x}=2t\sin(k_x).
    \label{eq:vhubbard}
\end{equation}
Observe that as $k\rightarrow 0$, $v_k\rightarrow k_x$ and if all momenta are small $F=0$.

We compute perturbatively to order $U^2$ so this amounts to evaluating the force-force correlator in the non-interacting ground state. 

In the dynamical mean field approximation to the Hubbard model the electron Green function is
\begin{equation}
G_{\sigma}(k,i\omega_n)=\frac{1}{i\omega_n+\mu-E_k-\Sigma_{\sigma}(i\omega_n)},
\label{eq:Gdmft}
\end{equation}

and the current-current correlation function is
\begin{widetext}
\bea
\chi_{jj}^{xx}(i\omega_n)&=& -T\sum_{\nu_n,\sigma}\sum_k v_kG_{\sigma}(k,i\omega_n+i\nu_n)v_kG_{\sigma}(k,i\nu_n) \nonumber \\
&\approx&-T\sum_{\nu_n,\sigma}\sum_k v_k^2\frac{\Sigma_{\sigma}(i\omega_n+i\nu_n)\left(i\nu_n+\mu-E_k\right)+\Sigma_{\sigma}(i\nu_n)\left(i\omega_n+i\nu_n+\mu-E_k\right)}{\left(i\omega_n+i\nu_n+\mu-E_k\right)^2\left(i\nu_n+\mu-E_k\right)^2}, \nonumber \\
\label{eq:chidmft}
\eea
\end{widetext}
where the second approximate equality holds to order $U^2$. In evaluating Eq.~(\ref{eq:chidmft}) we use the DMFT self energies computed as in Eq.~(\ref{eq:selfdmft}).

We will work in the paramagnetic phase of the model and will omit spin indices except where necessary.

\section{total spectral weight correction}
\label{sec:K}
We first look at the total spectral weight. From Eq.~(\ref{eq:KHubbard}), using $\langle c_{k\sigma}^{\dagger}c_{k\sigma}\rangle=G(k,\tau=0^{-})$ and  $G^{-1}(k,i\omega_n)=G_0^{-1}(k,i\omega_n)-\Sigma(k,i\omega_n)$ with $G_0^{-1}(k,i\omega_n)=i\omega_n-\varepsilon_k$ and $\varepsilon_k=E_k-\mu$ and noting that the frequency sum is absolutely convergent, we obtain
\bea
K&=&2T\sum_{\omega_n} \sum_k \frac{\partial^2 E_k}{\partial k_x^2} G(k,i\omega_n)\nonumber \\
&=&2T\sum_{\omega_n} \sum_k \frac{\partial^2 E_k}{\partial k_x^2}\left(G_0(k,i\omega_n)+\Sigma(k,i\omega_n)G_0^2(k,i\omega_n)\right) \nonumber \\
\label{eq:K2}
\eea
where the factor of $2$ comes from spin and $T$ represents temperature. $\Sigma\sim U^2$ at small $U$ so the second term of Eq.~(\ref{eq:K2}) gives the  spectral weight correction $\delta K$ to order $U^2$.  In performing the sum the double pole arising from the $G_0^2$ has to be handled with care. We find (see Appendix~\ref{app:total} for the details) that the exact perturbative result is
\bea
\delta K &=& U^2\sum_{kk'q}\frac{(1-f(\varepsilon_k))(1-f(\varepsilon_{k'}))f(\varepsilon_{k+q})f(\varepsilon_{k'-q})}{(\varepsilon_{k}+\varepsilon_{k'}-\varepsilon_{k+q}-\varepsilon_{k'-q})^2} \nonumber \\
&\times&(T_k+T_{k'}-T_{k+q}-T_{k'-q}),
\label{Kp}
\eea
and the DMFT approximation is
\bea
\delta K_{DMFT} &=& U^2\sum_{kk'qq'}\frac{(1-f(\varepsilon_k))(1-f(\varepsilon_{k'}))f(\varepsilon_{q})f(\varepsilon_{q'})}{(\varepsilon_{k}+\varepsilon_{k'}-\varepsilon_{q}-\varepsilon_{q'})^2} \nonumber \\
&\times&(T_k+T_{k'}-T_{q}-T_{q'}). 
\label{Kd}
\eea
where $\displaystyle{T_k=\frac{\partial^2 \varepsilon_k}{\partial k_x^2}}$ and $\displaystyle{f(\varepsilon_k)=\frac{1}{e^{\beta\varepsilon_k}+1}}$ is the Fermi function, here evaluated at $T=0$.

 Notice the similarity between Eq.~(\ref{Kd}) and Eq.~(\ref{Kp}). The only difference is that in the DMFT expression, the momentum conservation is relaxed so one has a sum over four independent momenta.

\begin{figure}[h!]
\centering
\includegraphics[width=1.0\columnwidth]{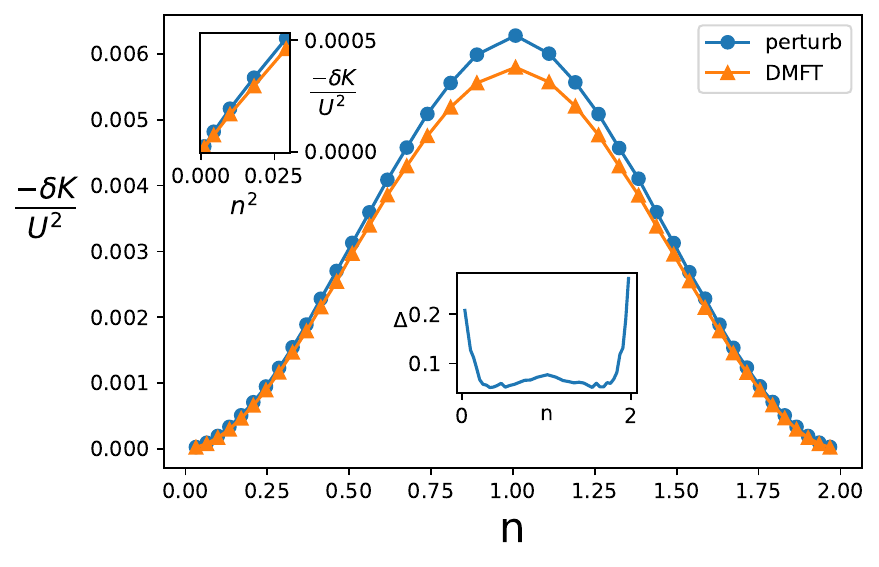}
\caption{Main panel: total spectral weight correction as a function of density obtained by numerical integration of Eqs.~(\ref{Kp}) and (\ref{Kd}). The upper left inset shows the low density dependence of total spectral weight correction as a function of $n^2$ and the inset in the middle shows the renormalized difference $\Delta$ between two curves which is defined as $\Delta=(\delta K_{perturb}-\delta K_{DMFT})/\delta K_{perturb}$.}
\label{fig:K}
\end{figure}

We have evaluated Eqs.~(\ref{Kp}) and (\ref{Kd}) numerically using Monte Carlo integration with $10^7$ points. Results are shown in Fig.~\ref{fig:K}. We can see that the total spectral weight correction is almost the same in the two methods. 

The very close correspondence  of the two results shows the accuracy of DMFT in calculating local expectation values even in the two dimensional case. Note in particular that in the very low density limit the two expressions are indistinguishable on the scale of the main panel, both vanishing $\sim n^2$ although with slightly different coefficients (see inset). This correspondence shows that the $n\rightarrow 0$ limit of the Hubbard model is not precisely a Galilean invariant theory. Although the dispersion for electrons near the Fermi level approaches $k^2$,  the interaction correction to the total kinetic energy scales in the same way as any other local interaction effect, namely $\sim n^2$. If the low density limit of the model were described by a theory that became truly Galilean-invariant in the senses described above,  we would expect the correction to vanish as a higher power of $n$. The reason that the interaction correction to the total spectral weight does not vanish more rapidly than $n^2$ may be seen in Fig.~\ref{fig:together}: interaction effects cause optical transitions at very high frequencies: the final states in these transitions are high up in the band where the dispersion is not well approximated by $k^2/2m$ and arguments based on Galilean invariance do not apply.

\section{Conductivity at nonzero frequency}
\label{sec:ffp}
We evaluate $\Gamma$ from Eq.~(\ref{eq:Mandchijj}) and then compute the conductivity from Eq.~(\ref{eq:sigmaasmemoryfunctionexpand}). Using Eq.~(\ref{eq:FHubbard}) for the force operator, we get (see Appendix~\ref{app:pert} for details) at zero temperature and positive frequency
\begin{widetext}
\bea
Re\sigma(\omega)=U^2\frac{\pi}{\omega^3}\sum_{kk'q}f(\varepsilon_k)f(\varepsilon_{k'})(1-f(\varepsilon_{k'-q}))(1-f(\varepsilon_{k+q})) (v_{k+q}+v_{k'-q}-v_{k}-v_{k'})^2\delta(\omega+\varepsilon_k+\varepsilon_{k'}-\varepsilon_{k'-q}-\varepsilon_{k+q}). 
\nonumber\\
\label{Kfp}
\eea
\end{widetext}

For the purposes of numerical evaluation it is convenient to rewrite Eq.~(\ref{Kfp}) as 
\bea
Re\sigma(\Omega)&=&\frac{2U^2}{\pi\Omega^3}\sum_{q}\int_{-\Omega}^0 d\omega \bigg[B^{(2)}(q,\omega+\Omega)B^{(0)}(-q,-\omega) \nonumber \\
&+&B^{(1)}(q,\omega+\Omega)B^{(1)}(-q,-\omega)\bigg]. 
\eea
where 
\bea
B^{\alpha}(q,\omega)&=&-\pi\sum_{k}(f(\varepsilon_k)-f(\varepsilon_{k+q}))\delta(\omega+\varepsilon_k-\varepsilon_{k+q})(v_{k+q}-v_{k})^\alpha, \nonumber \\
\alpha&=&0,1,2.
\eea

We evaluate $B^{\alpha}(q,\omega)$ by analytically implementing the delta function and performing the remaining integral via a standard trapezoid rule. Then we calculate the convolution to obtain the conductivity.

The DMFT conductivity is obtained from Eq.~(\ref{eq:chidmft}). Continuing to real frequency we obtain
\begin{widetext}
\bea
Im\chi_{jj}^{xx}(\omega)&=&2\sum_k\big(\frac{\partial \varepsilon_k}{\partial k_x}\big)^2\int \frac{dy}{\pi} (f(y)-f(y+\omega))\frac{\Sigma''(y+\omega)}{(y+\omega-\varepsilon_k-\Sigma'(y+w))^2+(\Sigma''(y+\omega))^2} \nonumber \\
&\times& \frac{\Sigma''(y)}{(y-\varepsilon_k-\Sigma'(y))^2+(\Sigma''(y))^2}.
\eea
\end{widetext}

We assume that the imaginary part of the self energy is small so we approximate the two Lorentzian above as Dirac delta function. After  calculations (details in Appendix~\ref{app:dmft}), we get at zero temperature and positive frequency,
\begin{widetext}
\bea
Re\sigma(\omega)=U^2\frac{\pi}{\omega^3}\sum_{kk'p_1p_2}f(\varepsilon_k)f(\varepsilon_{k'})(1-f(\varepsilon_{p_1}))(1-f(\varepsilon_{p_2}))\Big[v_k^2+v_{k'}^2+v_{p_1}^2+v_{p_2}^2\Big]\delta(\varepsilon_k+\varepsilon_{k'}+\omega-\varepsilon_{p_1}-\varepsilon_{p_2}).
\label{Kfd}
\eea
\end{widetext}

As in the expression for $\delta K$, the perturbative and DMFT approximation differ only by a relaxation of  momentum conservation, which causes the cross terms in the matrix element to vanish in the DMFT expression.

\begin{figure}[h!]
\centering
\includegraphics[width=1.0\columnwidth]{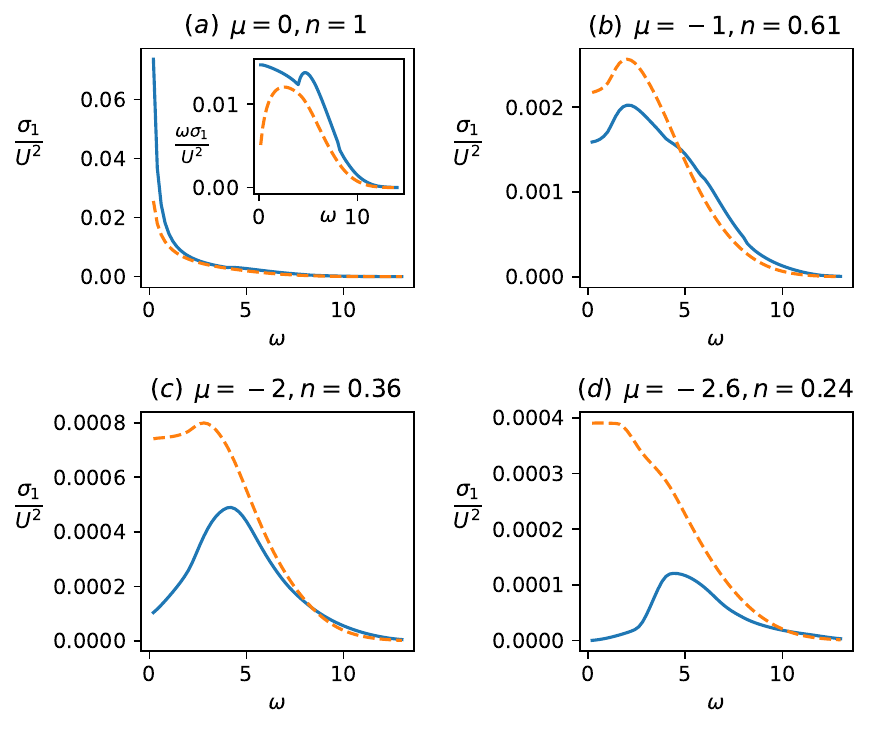}
\caption{Optical conductivity for different chemical potential for the perturbative (solid blue) and DMFT (dashed yellow)  cases ($\omega>0$). The inset in \text{(a)} shows the conductivity for half filled case multiplied by frequency.}
\label{fig:together}
\end{figure}

Fig.~\ref{fig:together} presents a detailed comparison between the perturbative and DMFT results for the conductivity. At high frequency the two methods give almost identical conductivities while at low frequency differences are evident. The differences are larger for lower $\mu$, becoming qualitative for $\mu<-2t$. At $\mu=0$ a qualitatively different low frequency behavior is also evident. 

\begin{figure*}
\includegraphics[scale=0.6]{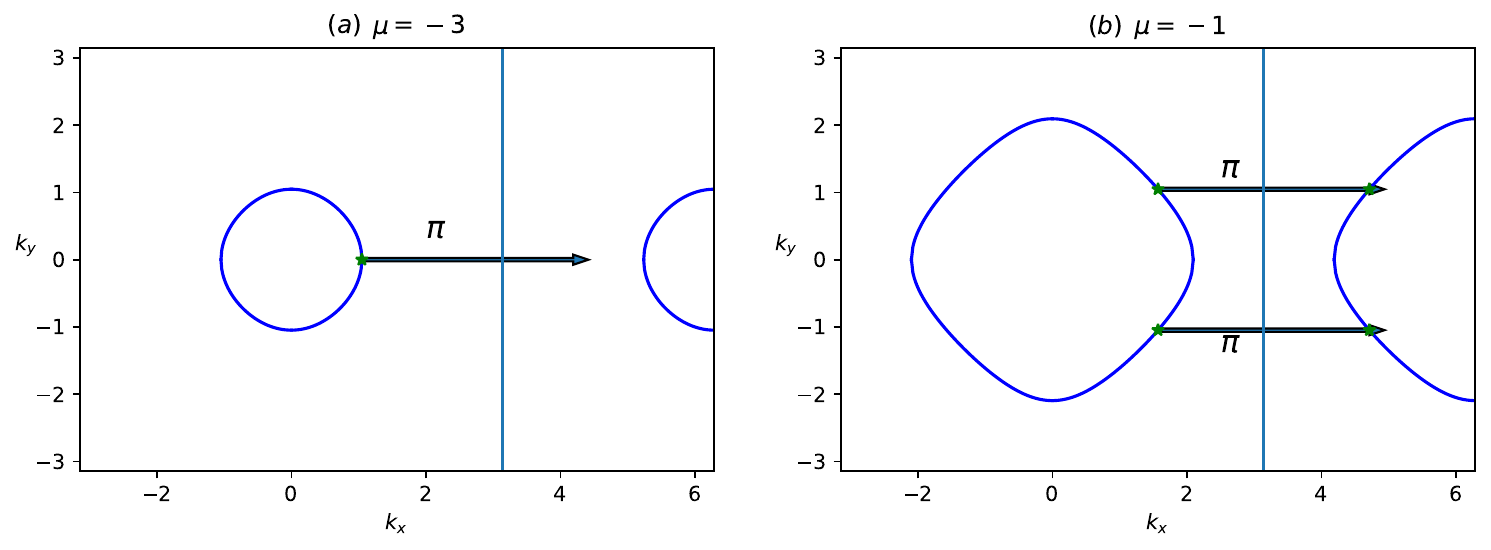}
\caption{(a) When the chemical potential $\mu<-2t$, there's no Umklapp scattering at low frequency. (b) When $-2t<\mu<2t$ Umklapp scattering is allowed at low frequency.}
\label{fig:scatter}
\end{figure*}

First focus on the exact perturbative case. We can see that when the chemical potential is larger than $-2t$ (or equivalently $k_F>G/4$, here $G$ is a reciprocal lattice vector) as in panel (b) the conductivity tends to  a non-zero constant as $\omega$ goes to zero. When the chemical potential is smaller than $-2t$ as in panel (d) the  conductivity vanishes at low frequency, with the first correction $\sim \omega^2$. This behavior was previously noted by Rosch and Howell \cite{rosch2005zero}, who showed that at small chemical potential only cooper channel scattering (from $k,-k$ to $q,-q$) is allowed. This zero crystal momentum process cannot degrade the long time limit of the current but because the current is not equivalent to the momentum in a lattice model, the current will have time dependence at shorter times, leading to the $\omega^2$ behavior. %,  is small there's no umklapp scattering (basically the electrons can only scatter within the first Brillouin zone) while when the chemical potential is greater than $-2t$ the umklapp process can occur (basically electrons can be scattered off to the next zone), as shown in Fig.~\ref{fig:scatter}. If there's no umklapp scattering then in the zero frequency limit the optical conductivity approaches zero because only cooper channel scattering (from $k,-k$ to $q,-q$) is allowed. This is a zero crystal momentum process and conserves the current. 
However for $-2t<\mu<2t$ Umklapp scattering processes in which after translation by a reciprocal lattice vector $G$ a pair of electrons can scatter across the Fermi surface may occur (see Fig.~\ref{fig:scatter}); these processes change the momentum of the system, resulting in a nonzero conductivity in the zero frequency limit. Notice that these are the results of order $U^2$. If we go to order $U^4$ which involves four particle processes, then the threshold for Umklapp scattering is much smaller.
%\begin{figure}[htbp!]
%\includegraphics[width=0.9\columnwidth]{scatter.pdf}
%\includegraphics[scale=0.8]{scatter.pdf}
%\caption{(a) When the chemical potential $\mu<-2t$, there's no Umklapp scattering at low frequency. (b) When $-2t<\mu<2t$ Umklapp scattering is allowed at low frequency.}
%\label{fig:scatter}
%\end{figure}

In the DMFT case, we can see that at all $\mu$ the conductivity remains nonzero  as $\omega$ goes to zero  except around the half filled case. This is due to the lack of vertex corrections in DMFT, in other words,  in DMFT all scattering can relax the current:  the DMFT calculation does not capture the effective Galilean invariance emerging at low densities. 

\begin{figure}[h!]
\centering
\includegraphics[width=1.0\columnwidth]{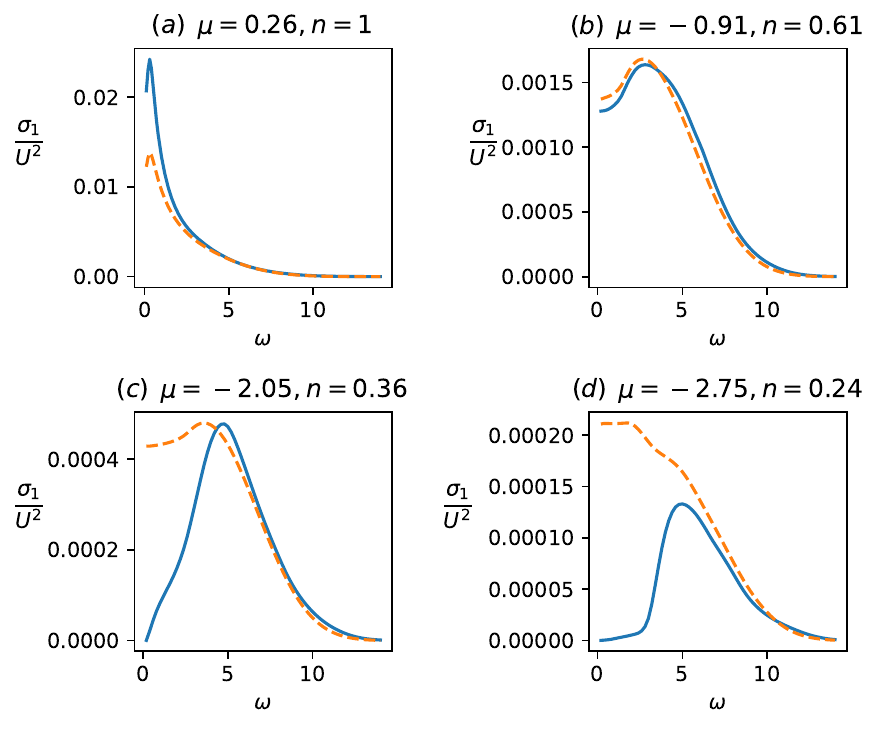}
\caption{Optical conductivity for the same carrier densities as in Fig.~\ref{fig:together} but with $t'=0.1$ for the perturbative (solid blue) and DMFT (dashed yellow) cases ($\omega>0$).}
\label{fig:halftp}
\end{figure}

Now we look at the $\mu=0$ case which corresponds to half filling. To better characterize the behaviour of the conductivity at low frequency we plot $\omega\sigma(\omega)$ in the inset of Fig.~\ref{fig:together}. We can see that for the perturbative case $\omega\sigma$ is weakly increasing as $\omega\rightarrow 0$. This divergence arises from the Van Hove and nesting properties of the nearest neighbor Hubbard model at $n=1$ and $t^\prime=0$, and is cut off at low frequency by the spin density wave gap  that arises from the nesting. At frequencies less than the gap scale the real part of the conductivity would vanish. We do not explicitly consider this physics; our results are valid only at frequencies sufficiently greater than the gap scale, which is  exponentially small in $\sqrt{t/U}$. %To be more concrete, the gap below which the conductivity vanishes is of the order of $e^{-\frac{1}{\sqrt{U}}}$ because of both nesting and Van Hove. The optical conductivity is roughly $U^2\frac{1}{\omega}$ from this gap up to some cutoff $w^\star$ and the integral will give $U^{\frac{3}{2}}$. Thus we obtain an order $U^\frac{3}{2}$ quantity which matches our expectation for the perturbative expansion.
At half filling DMFT disagrees with the perturbative calculation because  the momentum average wipes out the effects arising from nesting and Van Hove singularities, leading to an underestimate of the scattering rate. % because it bends down already at a large enough frequency for the $\omega\sigma(\omega)$. The exact perturbative $\omega\sigma(\omega)$ result would give a almost constant (more accurately speaking slightly increasing from the plot) down to some frequency of the order of the gap $\Delta\sim e^{-\frac{1}{\sqrt{U}}}$ which is very small.

To further document the origin of the effect, we add a next nearest neighbour hopping $t'$ to our tight binding model so  the dispersion relation becomes $E_k=-2t\cos(k_x)-2t\cos(k_y)-4t'\cos(k_x)\cos(k_y)$ and at half filling the perfect nesting is destroyed and the energy of the Van hove point is shifted.  Fig.~\ref{fig:halftp} shows the optical conductivity for this case for the same carrier densities as Fig.~\ref{fig:together}. We can see that now for half filling at low frequency the conductivity approaches a nonzero constant.

\section{Drude Weight}
\label{sec:dw}
\begin{figure}[h!]
\centering
\includegraphics[width=1.0\columnwidth]{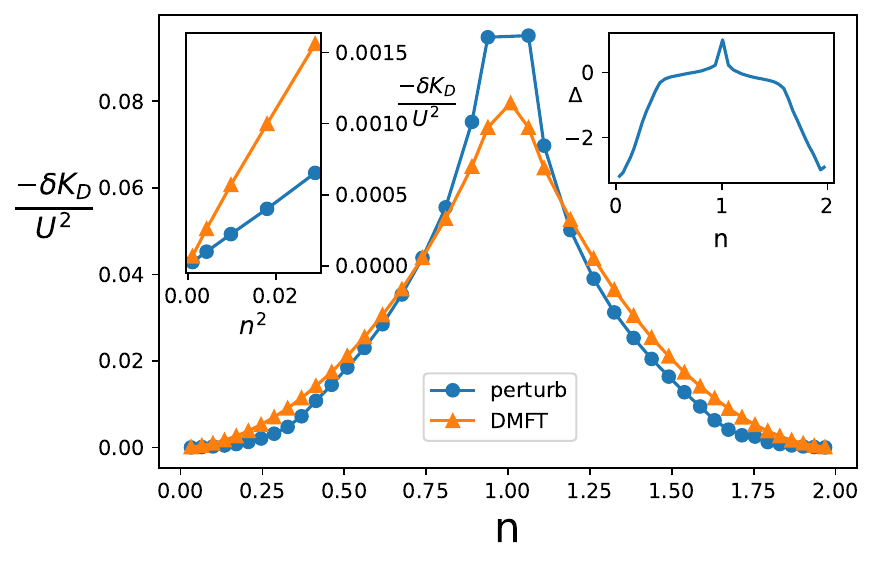}
\caption{Main panel: Drude weight correction as a function of density. The upper left inset shows the low density dependence of Drude weight correction as a function of $n^2$ and the upper right inset shows the renormalized difference  between two curves   $\Delta=(\delta K_{D_{perturb}}-\delta K_{D_{DMFT}})/\delta K_{D_{perturb}}$. Note that in the exact calculation $-\delta K_{D_{perturb}}/U^2$ diverges as $n\rightarrow 1$.}
\label{fig:drude}
\end{figure}

Fig.~\ref{fig:drude} shows the  Drude weight $K_D$ defined in Eq.~(\ref{eq:sigmalow}). $K_D$ characterizes the fraction of the carriers that are freely accelerated by an electric field at $T=0$.  For most of the carrier concentration range the differences between the DMFT and exact perturbative calculations are not large, but two features of the results are noteworthy.

Near half filling the suppression of the Drude weight is greater in the exact perturbative calculation than in the DMFT calculation. This may be understood as a precursor of the spin density wave state. In the exact perturbative calculation the spectral weight in the non-zero  frequency calculation diverges at least logarithmically (as follows from the $\approx \omega^{-1}$ divergence of the conductivity shown in Fig.~\ref{fig:together}), implying an infinite renormalization of the Drude weight. The divergence is cut off by the spin density wave gap which is itself exponentially small in $\sqrt{t/U}$.  

Near the empty band the suppression of the Drude weight is much less in the exact perturbative calculation than in the DMFT calculation. Close comparison of the insets to Figs.~\ref{fig:K} and \ref{fig:drude} shows that while in the DMFT calculation the change to $K_D$ is about three times the change in $K$, in the exact perturbative calculation the change to $K_D$ is only about $20\%$ larger than the change in $K$. In Fermi liquid theory one may write the Drude weight as  $K_D=K(U)\frac{m}{m^*}(1+F_{1S}/2)$ , where $F_{1S}$ is the spin-symmetric Landau parameter with angular momentum $L=1$  \cite{Negele98,Abrikosov63}. In a two-dimensional Galilean-invariant system $(1+F_{1S}/2)=\frac{m^\star}{m}$ consistent with the statement that $K_D=K$. Our result for $K_D/K$  then suggests that the effective low energy theory describing the low energy physics  while not quite Galilean-invariant, is close to being so. In this sense the model develops an emergent approximate Galilean invariance. % This is a sign of an emergent Galilean invariance where the vertex correction acts to cancel the mass enhancement.In a two dimensional Fermi liquid the Drude weight may be written

 %However we note that the cancellation is not complete and the scaling of the correction is the same in the exact and DMFT calculations. Thus in the exact calculation we can conclude that the Landau parameter $F_{1S}/2\neq \frac{m^*}{m}-1$

\section{Conclusion}
\label{sec:conclusion}
The single site dynamical mean field approximation is based on a severe locality approximation; in view of the considerable success of the method it is of interest to critically examine the accuracy of the approximation. In this paper we have considered the question in the context of the optical conductivity, a response function which in certain cases--in particular for a  Galilean-invariant or nearly Galilean invariant situation is crucially controlled by non-locality, and for the two dimensional Hubbard model, far from the limit of infinite dimensionality where dynamical mean field theory is strictly valid. Our analysis was based on the availability of exact perturbative results for the frequency dependence of the $T=0$ conductivity.  

We found that over relatively wide ranges of density the dynamical mean field approximation gives semiquantitatively accurate (within $\sim 20\%$) results. The two exceptions are very near to half filling in the perfectly nested model, where the momentum averaging inherent in the DMFT approximation leads to an underestimate of the effects of the nesting on the electron scattering, and at relatively low densities where an approximate kind of Galilean invariance emerges, leading to a substantial difference between the scattering processes that give an electron lifetime and the scattering processes that can degrade a current. We observe that in general in a lattice model such as the Hubbard model, high frequency scattering processes are sensitive to the lattice and can degrade the current, but at low density, low frequency processes cannot degrade the current (see Figs.~\ref{fig:together} and \ref{fig:scatter}). For this reason at low densities the exact low frequency conductivity differs substantially from the predictions of the DMFT approximation, in particular vanishing as frequency tends to zero. This behavior may be viewed as an effective Galilean invariance of the low energy theory, at low dopings.  The presence of high frequency conductivity, however, means that even in the low density limit, the Landau parameter $F_{1S}$, while significantly different from zero, is not quite equal to $m^*/m$.

The results presented here were obtained in the weak interaction limit at $T=0$ of a two dimensional model. Extension to three dimension, to stronger interactions and to systems near critical points (see e.g. \cite{Caprara07}) would be of interest. An important open question is the temperature dependence of the resistivity. The methods introduced here are not directly applicable at $T\neq 0$, but the results do imply constraints on a $T>0$ theory, and it is an important open question question whether the force force correlation function considered here can be resummed to obtain a theory of the temperature dependence.

\begin{acknowledgments}  A. M., Z. S. and A. J. M. acknowledge support from the US Department of Energy (DOE), Office of Science, Basic Energy Sciences (BES), under award no. DE- SC0018426.
\end{acknowledgments}

\appendix
\begin{widetext}
\section{Total spectral weight}\label{app:total}
In this section we present the details of the derivation of Eqs.~(\ref{Kp}), (\ref{Kd}) of the main text.  Evaluating the self energies as described below we find that Eq.~(\ref{eq:K2}) may be written for both the perturbative and DMFT cases as
\begin{equation}
    \delta K=2U^2\sum_{k,p_1,p_2,p_3}T\sum_{\omega_n}\frac{\frac{\partial^2 \varepsilon_k}{\partial k_x^2}S_{k,p_1,p_2,p_3}}{(i\omega_n-\varepsilon_k)^2(i\omega_n+\varepsilon_{p_3}-\varepsilon_{p_1}-\varepsilon_{p_2})},
    \label{eq:deltaK1}
    \end{equation}
where $S$ is a combination of Fermi functions and (in the perturbative case) momentum conserving $\delta$ functions.

Evaluating the Matsubara sum in the standard way,   taking into account the double pole gives
\begin{equation}
    \delta K=2U^2\sum_{k,p_1,p_2,p_3}\frac{\frac{\partial^2 \varepsilon_k}{\partial k_x^2}S_{k,p_1,p_2,p_3}(f(\varepsilon_{p_1}+\varepsilon_{p_2}-\varepsilon_{p_3})-f(\varepsilon_k))}{(\varepsilon_k+\varepsilon_{p_3}-\varepsilon_{p_1}-\varepsilon_{p_2})^2}+ 2\sum_k\frac{\partial^2\varepsilon_k}{\partial k_x^2} Re\left[\Sigma(k,\varepsilon_k)\right]\frac{df(z)}{dz}\Bigr\rvert_{z=\varepsilon_k}.
    \label{eq:deltaK2}
    \end{equation}
    The second term may be interpreted as the leading correction to the difference between the kinetic energy evaluated in a noninteracting theory using the bare Fermi surface $E_k=\mu$ and using the renormalized Fermi surface $E_k+Re\Sigma(k,0)=\mu$. To see this more concretely, note that the square lattice symmetry means we may replace $\displaystyle{\frac{\partial^2\varepsilon_k}{\partial k_x^2}}$ by $\displaystyle{\frac{1}{2}(\frac{\partial^2\varepsilon_k}{\partial k_x^2}+\frac{\partial^2\varepsilon_k}{\partial k_y^2})=-\frac{1}{2}E_k=-\frac{\mu}{2}}$ where the last equality follows from the $T\rightarrow 0$ limit of $df/dz$. The term is then recognized as the order $U^2$ change in particle density $n$ if the calculation is performed at fixed $\mu$. If the calculation is instead performed at fixed particle density the chemical potential must be adjusted in a way that compensates for this term., which we will ignore henceforth.
    
We now turn to the other factors. The self energy 
$\Sigma$ is given by the standard convolution of three bare Green functions,
\bea
\Sigma_{\sigma}(k,i\Omega_n)=-U^2T^2\sum_{\omega_1\omega_2}\sum_{p_1}\sum_{p_2}G^{0}_{-\sigma}(p_1,i\omega_1)G^0_{\sigma}(p_2,i\omega_2)G^0_{-\sigma}(p_1+p_2-k,i\omega_1+i\omega_2-i\Omega_n).
\label{sigmau2} \nonumber \\
\eea
Evaluating in the standard way we obtain
\bea\Sigma(k,i\omega_n)&=&U^2\sum_{p_1p_2}\frac{f(\varepsilon_{p_1+p_2-k})(1-f(\varepsilon_{p_1}))(1-f(\varepsilon_{p_2}))+(1-f(\varepsilon_{p_1+p_2-k}))f(\varepsilon_{p_1})f(\varepsilon_{p_2})}{\varepsilon_{p_1+p_2-k}+i\omega_n-\varepsilon_{p_1}-\varepsilon_{p_2}}, \nonumber \label{eq:selfp}\eea
so that 
\begin{equation}
    S_{k,p_1,p_2,p_3}=\left(f(\varepsilon_{p_3})(1-f(\varepsilon_{p_1}))(1-f(\varepsilon_{p_2}))+(1-f(\varepsilon_{p_3}))f(\varepsilon_{p_1})f(\varepsilon_{p_2})\right)\delta_{p_3+k-p_1-p_2}
\end{equation}
for the perturbative case.

Combining the Fermi functions using equations such as
\bea
f(\varepsilon_{p_1+p_2-k})(1-f(\varepsilon_{p_2}))(1-f(\varepsilon_{p_1}))+(1-f(\varepsilon_{p_1+p_2-k}))f(\varepsilon_{p_1})f(\varepsilon_{p_2})=\frac{e^{\beta\varepsilon_{p_1+p_2-k}}(1+e^{\beta(\varepsilon_{p_1}+\varepsilon_{p_2}-\varepsilon_{p_1+p_2-k})})}{(1+e^{\beta\varepsilon_{p_1+p_2-k}})(1+e^{\beta\varepsilon_{p_1}})(1+e^{\beta\varepsilon_{p_2}})} \nonumber
\eea
and rearranging, this gives Eq.~(\ref{Kp}).

In the DMFT approximation, we have (correct only when interaction is weak)
\bea
\Sigma(i\omega_n)&=&\sum_k\Sigma(k,i\omega_n) \nonumber \\
&=& U^2\sum_{p_1p_2p_3}\frac{f(\varepsilon_{p_3})(1-f(\varepsilon_{p_1}))(1-f(\varepsilon_{p_2}))+(1-f(\varepsilon_{p_3}))f(\varepsilon_{p_1})f(\varepsilon_{p_2})}{\varepsilon_{p_3}+i\omega_n-\varepsilon_{p_1}-\varepsilon_{p_2}},
\label{eq:selfdmft}
\eea
so that
\bea
    S_{k,p_1,p_2,p_3}=f(\varepsilon_{p_3})(1-f(\varepsilon_{p_1}))(1-f(\varepsilon_{p_2}))+(1-f(\varepsilon_{p_3}))f(\varepsilon_{p_1})f(\varepsilon_{p_2}).
\eea
Similar steps give Eq.~(\ref{Kd}).

\section{Perturbative optical conductivity}\label{app:pert}
The current-current correlation function is
\bea
\chi_{jj}^{xx}(i\omega_n)&=&\int_{0}^{\beta}d\tau e^{iw_n\tau}\langle T_{\tau}\hat{j}^x(\tau)\hat{j}^x(0)\rangle \nonumber \\
&=&\frac{1}{i\omega_n}\langle \hat{j}^x(\beta)\hat{j}^x(0)-\hat{j}^x(0)\hat{j}^x(0)\rangle-\frac{1}{i\omega_n}\int_0^{\beta}d\tau e^{i\omega_n\tau}\langle T_{\tau}\frac{d\hat{j}^x(\tau)}{d\tau}j^x(0)\rangle. \nonumber 
\eea
The first term is zero because it's equal to $-\langle[\hat{j}^x,\hat{j}^x]\rangle$ and is the commutator of current with itself. Using time translational invariance we have
\bea
{\langle T_{\tau}\frac{d\hat{j}^x(\tau)}{d\tau}\hat{j}^x(0)\rangle=\langle T_{\tau}\frac{d\hat{j}^x(\tau')}{d\tau'}\bigg|_{\tau'=0}\hat{j}^x(-\tau)\rangle}. \nonumber
\eea

Now integrate by parts,
\bea
\chi_{jj}^{xx}(i\omega_n)&=&-\frac{1}{(i\omega_n)^2}\langle [\hat{j}^x(0),\frac{d\hat{j}^x(\tau)}{d\tau}\bigg|_{\tau=0}]\rangle+\frac{1}{(i\omega_n)^2}\int_0^{\beta}e^{i\omega_n\tau}\langle T_{\tau}\frac{d\hat{j}^x(\tau')}{d\tau'}\bigg|_{\tau'=0}\frac{d\hat{j}^x(-\tau)}{d\tau}\rangle \nonumber \\
&=& -\frac{1}{(i\omega_n)^2}\int_{0}^{\beta}d\tau e^{i\omega_n\tau}\langle T_{\tau}\frac{d\hat{j}^x(\tau)}{d\tau}\frac{d\hat{j}^x(\tau')}{d\tau'}\bigg|_{\tau'=0}\rangle.
\label{eq:ordering}
\eea
The term which contains the commutator of $\hat{j}^x(0)$ and $\displaystyle{\frac{d\hat{j}^x(\tau)}{d\tau}}$ is just a constant and doesn't contribute to the imaginary part of $\chi_{jj}^{xx}(i\omega_n)$. 

Evaluating the time-ordering product (keeping to second order) in Eq.~(\ref{eq:ordering}) we get
%\begin{multline*}
\bea
&-&\sum_{kk'q}\sum_{mnp}\int_{0}^{\beta}d\tau e^{i\omega_n\tau}\langle T_{\tau}\wick{\c1 c_{k\uparrow}^{\dagger}(\tau) \c2 c_{k'\downarrow}^{\dagger}(\tau) \c3 c_{k'-q\downarrow}(\tau)\c4 c_{k+q\uparrow}(\tau)\c4 c_{m\uparrow}^{\dagger}(0)\c3 c_{n\downarrow}^{\dagger}(0) \c2 c_{n-p\downarrow}(0)\c1 c_{m+p\uparrow}(0)}\rangle^{}_{0} \nonumber \\
&\times& (T_{k+q}+T_{k'-q}-T_{k}-T_{k'})(T_{m+p}+T_{n-p}-T_{m}-T_{n})  \nonumber \\
&=&\int_{0}^{\beta}d\tau e^{i\omega_n\tau}\sum_{kk'q}G_{\uparrow}^0(k,-\tau)G_{\downarrow}^0(k',-\tau)G_{\downarrow}^0(k'-q,\tau)G_{\uparrow}^{0}(k+q,\tau)(T_{k+q}+T_{k'-q}-T_{k}-T_{k'})^2 \nonumber \\
&=&\sum_{kk'q}\int_{0}^{\beta}d\tau e^{i\omega_n\tau}e^{\varepsilon_k\tau}e^{\varepsilon_{k'}\tau}e^{-\varepsilon_{k'-q}\tau}e^{-\varepsilon_{k+q}\tau}f(\varepsilon_k)f(\varepsilon_k')(1-f(\varepsilon_{k'-q}))(1-f(\varepsilon_{k+q})) \nonumber \\
&\times& (T_{k+q}+T_{k'-q}-T_{k}-T_{k'})^2. \nonumber
%\end{multline*}
\eea

Performing the integral over imaginary time, we obtain
\bea
\chi_{jj}^{xx}(i\omega_n)&=&U^2\frac{1}{(iw_n)^2}\sum_{kk'q}\frac{e^{\beta(\varepsilon_{k}+\varepsilon_{k'}-\varepsilon_{k'-q}-\varepsilon_{k+q})}-1}{i\omega_n+\varepsilon_k+\varepsilon_k'-\varepsilon_{k'-q}-\varepsilon_{k+q}}f(\varepsilon_k)f(\varepsilon_{k'})(1-f(\varepsilon_{k'-q}))(1-f(\varepsilon_{k+q})) \nonumber \\
&\times& (T_{k+q}+T_{k'-q}-T_{k}-T_{k'})^2.
\eea

Carrying out analytic continuation $iw_n\rightarrow w+i\delta$ and taking the imaginary part of $\chi_{jj}$, we get
\bea
\textit{Im}\chi_{jj}^{xx}(\omega)&=&U^2\frac{-\pi}{w^2}\sum_{kk'q}(e^{\beta(\varepsilon_k+\varepsilon_{k'}-\varepsilon_{k'-q}-\varepsilon_{k+q})}-1)f(\varepsilon_k)f(\varepsilon_{k'})(1-f(\varepsilon_{k'-q}))(1-f(\varepsilon_{k+q})) \nonumber \\
&\times &(T_{k+q}+T_{k'-q}-T_{k}-T_{k'})^2\delta(\omega+\varepsilon_k+\varepsilon_{k'}-\varepsilon_{k'-q}-\varepsilon_{k+q}).
\eea

Using the relation $\displaystyle{Re\sigma(\omega)=\frac{Im\chi_{jj}^{xx}(\omega)}{\omega}}$ at finite frequency, we have 
\bea
Re\sigma(\omega)&=&U^2\frac{-\pi}{\omega^3}\sum_{kk'q}(e^{\beta(\varepsilon_k+\varepsilon_{k'}-\varepsilon_{k'-q}-\varepsilon_{k+q})}-1)f(\varepsilon_k)f(\varepsilon_{k'})(1-f(\varepsilon_{k'-q}))(1-f(\varepsilon_{k+q})) \nonumber \\
&\times &(T_{k+q}+T_{k'-q}-T_{k}-T_{k'})^2\delta(\omega+\varepsilon_k+\varepsilon_{k'}-\varepsilon_{k'-q}-\varepsilon_{k+q}).
\eea

Then take the zero temperature limit and focus on positive frequency, we get Eq.~(\ref{Kfp}). Diagrammatically we are just evaluating the convolution of two bubbles, as shown in Fig.~\ref{fig:bubble}.
\begin{figure}[h!]
\centering
\includegraphics[scale=0.5]{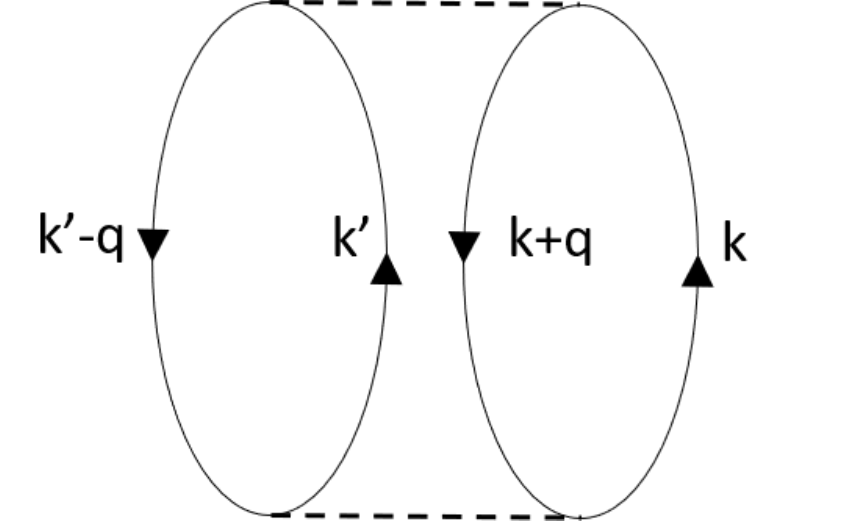}
\caption{Convolution of two bubbles.}
\label{fig:bubble}
\end{figure}

\section{Optical conductivity under DMFT}\label{app:dmft}
Using the spectral weight representation, we have
\bea
\chi_{jj}^{xx}(i\omega_n) &=& -2\sum_k \big(\frac{\partial \varepsilon_k}{\partial k_x}\big)^2T\sum_{\nu_n}\int \frac{dx}{\pi} \frac{ImG(k,x)}{i\omega_n+i\nu_n-x}\int \frac{dy}{\pi} \frac{ImG(k,y)}{i\nu_n-y}\nonumber \\
&=&-2\sum_k\big(\frac{\partial \varepsilon_k}{\partial k_x}\big)^2\int \frac{dx}{\pi} \frac{dy}{\pi} \frac{ImG(k,x)ImG(k,y)}{i\omega_n+y-x}(f(y)-f(x)). \nonumber \\
\eea

Then we take the imaginary part,
\bea
Im\chi_{jj}^{xx}(\omega) &=& 2\sum_k\big(\frac{\partial \varepsilon_k}{\partial k_x}\big)^2\int dx\frac{dy}{\pi}ImG(k,x)ImG(k,y)(f(y)-f(x))\delta(\omega+y-x) \nonumber \\
&=&2\sum_k\big(\frac{\partial \varepsilon_k}{\partial k_x}\big)^2\int \frac{dy}{\pi}ImG(k,y+\omega)ImG(k,y)(f(y)-f(y+\omega))\nonumber \\
&=&2\sum_k\big(\frac{\partial \varepsilon_k}{\partial k_x}\big)^2\int \frac{dy}{\pi} (f(y)-f(y+\omega))\frac{\Sigma''(y+\omega)}{(y+\omega-\varepsilon_k-\Sigma'(y+\omega))^2+(\Sigma''(y+\omega))^2} \nonumber \\
&\times& \frac{\Sigma''(y)}{(y-\varepsilon_k-\Sigma'(y))^2+(\Sigma''(y))^2}.
\eea

We introduce $\displaystyle{I(x)=\sum_k \delta(x-\varepsilon_k) \big(\frac{\partial \varepsilon_k}{\partial k_x}\big)^2}$ and approximate Lorentzian as Dirac delta functions. The real part of the conductivity is
\bea
Re\sigma(\omega)
&=& -2\int dy \int dx \frac{f(y)-f(y+\omega)}{\omega}I(x)\big[\delta(y+\omega-x)\frac{\Sigma''(y)}{\omega^2}+\frac{\Sigma''(y+\omega)}{\omega^2}\delta(y-x)\big] \nonumber \\
&=&\frac{-2}{\omega^3}\int dy(f(y)-f(y+\omega))[I(y+\omega)\Sigma''(y)+\Sigma''(y+\omega)I(y)].
\eea

Using the second order self energy Eq.~(\ref{eq:selfdmft}), we have
\bea
Re\sigma(\omega)&=&U^2\frac{2\pi}{\omega^3}\int dy (f(y)-f(y+\omega))\sum_{kk'p_1p_2} \delta(y+\omega-\varepsilon_k)\big(\frac{\partial \varepsilon_k}{\partial k_x}\big)^2\delta(\varepsilon_{k'}+y-\varepsilon_{p_1}-\varepsilon_{p_2}) \nonumber \\
&\times&[f(\varepsilon_{k'})(1-f(\varepsilon_{p_1}))(1-f(\varepsilon_{p_2}))+(1-f(\varepsilon_{k'}))f(\varepsilon_{p_1})f(\varepsilon_{p_2})] \nonumber \\
&+& U^2\frac{2\pi}{\omega^3}\int dy (f(y)-f(y+\omega))\sum_{kk'p_1p_2} \delta(y-\varepsilon_k)\big(\frac{\partial \varepsilon_k}{\partial k_x}\big)^2\delta(\varepsilon_k'+y+\omega-\varepsilon_{p_1}-\varepsilon_{p_2}) \nonumber \\
&\times&[f(\varepsilon_{k'})(1-f(\varepsilon_{p_1}))(1-f(\varepsilon_{p_2}))+(1-f(\varepsilon_{k'}))f(\varepsilon_{p_1})f(\varepsilon_{p_2})] \nonumber \\
&=&U^2\frac{2\pi}{\omega^3}\sum_{kk'p_1p_2}(f(\varepsilon_k-\omega)-f(\varepsilon_k))[f(\varepsilon_{k'})(1-f(\varepsilon_{p_1}))(1-f(\varepsilon_{p_2}))+(1-f(\varepsilon_{k'}))f(\varepsilon_{p_1})f(\varepsilon_{p_2})] \nonumber \\
&\times&\delta(\varepsilon_k+\varepsilon_{k'}-\omega-\varepsilon_{p_1}-\varepsilon_{p_2})\big(\frac{\partial \varepsilon_k}{\partial k_x}\big)^2\ \nonumber \\
&+&U^2\frac{2\pi}{\omega^3}\sum_{kk'p_1p_2}(f(\varepsilon_k)-f(\varepsilon_k+\omega))[f(\varepsilon_{k'})(1-f(\varepsilon_{p_1}))(1-f(\varepsilon_{p_2}))+(1-f(\varepsilon_{k'}))f(\varepsilon_{p_1})f(\varepsilon_{p_2})] \nonumber \\
&\times&\delta(\varepsilon_k+\varepsilon_{k'}+\omega-\varepsilon_{p_1}-\varepsilon_{p_2})\big(\frac{\partial \varepsilon_k}{\partial k_x}\big)^2.
\eea

Imposing the delta function, we get
\bea
Re\sigma(\omega)&=&U^2\frac{2\pi}{\omega^3}\sum_{kk'p_1p_2}[e^{\beta(\varepsilon_k+\varepsilon_k'-\varepsilon_{p_1}-\varepsilon_{p_2})}-1]f(\varepsilon_k)f(\varepsilon_{k'})(1-f(\varepsilon_{p_1}))(1-f(\varepsilon_{p_2})) \nonumber \\
&\times& \delta(\varepsilon_k+\varepsilon_{k'}-\omega-\varepsilon_{p_1}-\varepsilon_{p_2}) \big(\frac{\partial \varepsilon_k}{\partial k_x}\big)^2 \nonumber \\
&-&U^2\frac{2\pi}{\omega^3}\sum_{kk'p_1p_2}[e^{\beta(\varepsilon_k+\varepsilon_k'-\varepsilon_{p_1}-\varepsilon_{p_2})}-1]f(\varepsilon_k)f(\varepsilon_{k'})(1-f(\varepsilon_{p_1}))(1-f(\varepsilon_{p_2})) \nonumber \\
&\times&\delta(\varepsilon_k+\varepsilon_{k'}+\omega-\varepsilon_{p_1}-\varepsilon_{p_2}) \big(\frac{\partial \varepsilon_k}{\partial k_x}\big)^2 \nonumber \\
\eea

By changing variables we can combine these two terms above,
\bea
Re\sigma(\omega)&=&U^2\frac{-2\pi}{\omega^3}\sum_{kk'p_1p_2}[e^{\beta(\varepsilon_k+\varepsilon_k'-\varepsilon_{p_1}-\varepsilon_{p_2})}-1]f(\varepsilon_k)f(\varepsilon_{k'})(1-f(\varepsilon_{p_1}))(1-f(\varepsilon_{p_2})) \nonumber \\
&\times&\delta(\varepsilon_k+\varepsilon_{k'}+\omega-\varepsilon_{p_1}-\varepsilon_{p_2}) \bigg[\big(\frac{\partial \varepsilon_k}{\partial k_x}\big)^2+\big(\frac{\partial \varepsilon_{p_1}}{\partial p_{1_x}}\big)^2\bigg].
\eea

Taking the zero temperature limit and focus on positive frequency, we obtain Eq.~(\ref{Kfd}).

\end{widetext}

\bibliography{reference}

\begin{thebibliography}{16}%
\makeatletter
\providecommand \@ifxundefined [1]{%
 \@ifx{#1\undefined}
}%
\providecommand \@ifnum [1]{%
 \ifnum #1\expandafter \@firstoftwo
 \else \expandafter \@secondoftwo
 \fi
}%
\providecommand \@ifx [1]{%
 \ifx #1\expandafter \@firstoftwo
 \else \expandafter \@secondoftwo
 \fi
}%
\providecommand \natexlab [1]{#1}%
\providecommand \enquote  [1]{``#1''}%
\providecommand \bibnamefont  [1]{#1}%
\providecommand \bibfnamefont [1]{#1}%
\providecommand \citenamefont [1]{#1}%
\providecommand \href@noop [0]{\@secondoftwo}%
\providecommand \href [0]{\begingroup \@sanitize@url \@href}%
\providecommand \@href[1]{\@@startlink{#1}\@@href}%
\providecommand \@@href[1]{\endgroup#1\@@endlink}%
\providecommand \@sanitize@url [0]{\catcode `\\12\catcode `\$12\catcode
  `\&12\catcode `\#12\catcode `\^12\catcode `\_12\catcode `\%12\relax}%
\providecommand \@@startlink[1]{}%
\providecommand \@@endlink[0]{}%
\providecommand \url  [0]{\begingroup\@sanitize@url \@url }%
\providecommand \@url [1]{\endgroup\@href {#1}{\urlprefix }}%
\providecommand \urlprefix  [0]{URL }%
\providecommand \Eprint [0]{\href }%
\providecommand \doibase [0]{https://doi.org/}%
\providecommand \selectlanguage [0]{\@gobble}%
\providecommand \bibinfo  [0]{\@secondoftwo}%
\providecommand \bibfield  [0]{\@secondoftwo}%
\providecommand \translation [1]{[#1]}%
\providecommand \BibitemOpen [0]{}%
\providecommand \bibitemStop [0]{}%
\providecommand \bibitemNoStop [0]{.\EOS\space}%
\providecommand \EOS [0]{\spacefactor3000\relax}%
\providecommand \BibitemShut  [1]{\csname bibitem#1\endcsname}%
\let\auto@bib@innerbib\@empty
%</preamble>
\bibitem [{\citenamefont {Georges}\ \emph {et~al.}(1996)\citenamefont
  {Georges}, \citenamefont {Kotliar}, \citenamefont {Krauth},\ and\
  \citenamefont {Rozenberg}}]{georges1996dynamical}%
  \BibitemOpen
  \bibfield  {author} {\bibinfo {author} {\bibfnamefont {A.}~\bibnamefont
  {Georges}}, \bibinfo {author} {\bibfnamefont {G.}~\bibnamefont {Kotliar}},
  \bibinfo {author} {\bibfnamefont {W.}~\bibnamefont {Krauth}},\ and\ \bibinfo
  {author} {\bibfnamefont {M.~J.}\ \bibnamefont {Rozenberg}},\ }\bibfield
  {title} {\bibinfo {title} {Dynamical mean-field theory of strongly correlated
  fermion systems and the limit of infinite dimensions},\ }\href
  {https://doi.org/10.1103/RevModPhys.68.13} {\bibfield  {journal} {\bibinfo
  {journal} {Reviews of Modern Physics}\ }\textbf {\bibinfo {volume} {68}},\
  \bibinfo {pages} {13} (\bibinfo {year} {1996})}\BibitemShut {NoStop}%
\bibitem [{\citenamefont {Kotliar}\ \emph {et~al.}(2006)\citenamefont
  {Kotliar}, \citenamefont {Savrasov}, \citenamefont {Haule}, \citenamefont
  {Oudovenko}, \citenamefont {Parcollet},\ and\ \citenamefont
  {Marianetti}}]{Kotliar2006a}%
  \BibitemOpen
  \bibfield  {author} {\bibinfo {author} {\bibfnamefont {G.}~\bibnamefont
  {Kotliar}}, \bibinfo {author} {\bibfnamefont {S.~Y.}\ \bibnamefont
  {Savrasov}}, \bibinfo {author} {\bibfnamefont {K.}~\bibnamefont {Haule}},
  \bibinfo {author} {\bibfnamefont {V.~S.}\ \bibnamefont {Oudovenko}}, \bibinfo
  {author} {\bibfnamefont {O.}~\bibnamefont {Parcollet}},\ and\ \bibinfo
  {author} {\bibfnamefont {C.~A.}\ \bibnamefont {Marianetti}},\ }\bibfield
  {title} {\bibinfo {title} {{Electronic structure calculations with dynamical
  mean-field theory}},\ }\href {https://doi.org/10.1103/RevModPhys.78.865}
  {\bibfield  {journal} {\bibinfo  {journal} {Reviews of Modern Physics}\
  }\textbf {\bibinfo {volume} {78}},\ \bibinfo {pages} {865} (\bibinfo {year}
  {2006})}\BibitemShut {NoStop}%
\bibitem [{\citenamefont {Amadon}\ \emph {et~al.}(2008)\citenamefont {Amadon},
  \citenamefont {Lechermann}, \citenamefont {Georges}, \citenamefont {Jollet},
  \citenamefont {Wehling},\ and\ \citenamefont
  {Lichtenstein}}]{amadon2008plane}%
  \BibitemOpen
  \bibfield  {author} {\bibinfo {author} {\bibfnamefont {B.}~\bibnamefont
  {Amadon}}, \bibinfo {author} {\bibfnamefont {F.}~\bibnamefont {Lechermann}},
  \bibinfo {author} {\bibfnamefont {A.}~\bibnamefont {Georges}}, \bibinfo
  {author} {\bibfnamefont {F.}~\bibnamefont {Jollet}}, \bibinfo {author}
  {\bibfnamefont {T.~O.}\ \bibnamefont {Wehling}},\ and\ \bibinfo {author}
  {\bibfnamefont {A.~I.}\ \bibnamefont {Lichtenstein}},\ }\bibfield  {title}
  {\bibinfo {title} {Plane-wave based electronic structure calculations for
  correlated materials using dynamical mean-field theory and projected local
  orbitals},\ }\href {https://doi.org/10.1103/PhysRevB.77.205112} {\bibfield
  {journal} {\bibinfo  {journal} {Phys. Rev. B}\ }\textbf {\bibinfo {volume}
  {77}},\ \bibinfo {pages} {205112} (\bibinfo {year} {2008})}\BibitemShut
  {NoStop}%
\bibitem [{\citenamefont {Gull}\ \emph {et~al.}(2010)\citenamefont {Gull},
  \citenamefont {Ferrero}, \citenamefont {Parcollet}, \citenamefont {Georges},\
  and\ \citenamefont {Millis}}]{Gull10}%
  \BibitemOpen
  \bibfield  {author} {\bibinfo {author} {\bibfnamefont {E.}~\bibnamefont
  {Gull}}, \bibinfo {author} {\bibfnamefont {M.}~\bibnamefont {Ferrero}},
  \bibinfo {author} {\bibfnamefont {O.}~\bibnamefont {Parcollet}}, \bibinfo
  {author} {\bibfnamefont {A.}~\bibnamefont {Georges}},\ and\ \bibinfo {author}
  {\bibfnamefont {A.~J.}\ \bibnamefont {Millis}},\ }\bibfield  {title}
  {\bibinfo {title} {Momentum-space anisotropy and pseudogaps: A comparative
  cluster dynamical mean-field analysis of the doping-driven metal-insulator
  transition in the two-dimensional hubbard model},\ }\href
  {https://doi.org/10.1103/PhysRevB.82.155101} {\bibfield  {journal} {\bibinfo
  {journal} {Phys. Rev. B}\ }\textbf {\bibinfo {volume} {82}},\ \bibinfo
  {pages} {155101} (\bibinfo {year} {2010})}\BibitemShut {NoStop}%
\bibitem [{\citenamefont {LeBlanc}\ \emph {et~al.}(2015)\citenamefont
  {LeBlanc}, \citenamefont {Antipov}, \citenamefont {Becca}, \citenamefont
  {Bulik}, \citenamefont {Chan}, \citenamefont {Chung}, \citenamefont {Deng},
  \citenamefont {Ferrero}, \citenamefont {Henderson}, \citenamefont
  {Jim\'enez-Hoyos}, \citenamefont {Kozik}, \citenamefont {Liu}, \citenamefont
  {Millis}, \citenamefont {Prokof'ev}, \citenamefont {Qin}, \citenamefont
  {Scuseria}, \citenamefont {Shi}, \citenamefont {Svistunov}, \citenamefont
  {Tocchio}, \citenamefont {Tupitsyn}, \citenamefont {White}, \citenamefont
  {Zhang}, \citenamefont {Zheng}, \citenamefont {Zhu},\ and\ \citenamefont
  {Gull}}]{LeBlanc15}%
  \BibitemOpen
  \bibfield  {author} {\bibinfo {author} {\bibfnamefont {J.~P.~F.}\
  \bibnamefont {LeBlanc}}, \bibinfo {author} {\bibfnamefont {A.~E.}\
  \bibnamefont {Antipov}}, \bibinfo {author} {\bibfnamefont {F.}~\bibnamefont
  {Becca}}, \bibinfo {author} {\bibfnamefont {I.~W.}\ \bibnamefont {Bulik}},
  \bibinfo {author} {\bibfnamefont {G.~K.-L.}\ \bibnamefont {Chan}}, \bibinfo
  {author} {\bibfnamefont {C.-M.}\ \bibnamefont {Chung}}, \bibinfo {author}
  {\bibfnamefont {Y.}~\bibnamefont {Deng}}, \bibinfo {author} {\bibfnamefont
  {M.}~\bibnamefont {Ferrero}}, \bibinfo {author} {\bibfnamefont {T.~M.}\
  \bibnamefont {Henderson}}, \bibinfo {author} {\bibfnamefont {C.~A.}\
  \bibnamefont {Jim\'enez-Hoyos}}, \bibinfo {author} {\bibfnamefont
  {E.}~\bibnamefont {Kozik}}, \bibinfo {author} {\bibfnamefont {X.-W.}\
  \bibnamefont {Liu}}, \bibinfo {author} {\bibfnamefont {A.~J.}\ \bibnamefont
  {Millis}}, \bibinfo {author} {\bibfnamefont {N.~V.}\ \bibnamefont
  {Prokof'ev}}, \bibinfo {author} {\bibfnamefont {M.}~\bibnamefont {Qin}},
  \bibinfo {author} {\bibfnamefont {G.~E.}\ \bibnamefont {Scuseria}}, \bibinfo
  {author} {\bibfnamefont {H.}~\bibnamefont {Shi}}, \bibinfo {author}
  {\bibfnamefont {B.~V.}\ \bibnamefont {Svistunov}}, \bibinfo {author}
  {\bibfnamefont {L.~F.}\ \bibnamefont {Tocchio}}, \bibinfo {author}
  {\bibfnamefont {I.~S.}\ \bibnamefont {Tupitsyn}}, \bibinfo {author}
  {\bibfnamefont {S.~R.}\ \bibnamefont {White}}, \bibinfo {author}
  {\bibfnamefont {S.}~\bibnamefont {Zhang}}, \bibinfo {author} {\bibfnamefont
  {B.-X.}\ \bibnamefont {Zheng}}, \bibinfo {author} {\bibfnamefont
  {Z.}~\bibnamefont {Zhu}},\ and\ \bibinfo {author} {\bibfnamefont
  {E.}~\bibnamefont {Gull}} (\bibinfo {collaboration} {Simons Collaboration on
  the Many-Electron Problem}),\ }\bibfield  {title} {\bibinfo {title}
  {Solutions of the two-dimensional hubbard model: Benchmarks and results from
  a wide range of numerical algorithms},\ }\href
  {https://doi.org/10.1103/PhysRevX.5.041041} {\bibfield  {journal} {\bibinfo
  {journal} {Phys. Rev. X}\ }\textbf {\bibinfo {volume} {5}},\ \bibinfo {pages}
  {041041} (\bibinfo {year} {2015})}\BibitemShut {NoStop}%
\bibitem [{\citenamefont {Wietek}\ \emph {et~al.}(2021)\citenamefont {Wietek},
  \citenamefont {Rossi}, \citenamefont {\ifmmode~\check{S}\else
  \v{S}\fi{}imkovic}, \citenamefont {Klett}, \citenamefont {Hansmann},
  \citenamefont {Ferrero}, \citenamefont {Stoudenmire}, \citenamefont
  {Sch\"afer},\ and\ \citenamefont {Georges}}]{Wietek21}%
  \BibitemOpen
  \bibfield  {author} {\bibinfo {author} {\bibfnamefont {A.}~\bibnamefont
  {Wietek}}, \bibinfo {author} {\bibfnamefont {R.}~\bibnamefont {Rossi}},
  \bibinfo {author} {\bibfnamefont {F.}~\bibnamefont {\ifmmode~\check{S}\else
  \v{S}\fi{}imkovic}}, \bibinfo {author} {\bibfnamefont {M.}~\bibnamefont
  {Klett}}, \bibinfo {author} {\bibfnamefont {P.}~\bibnamefont {Hansmann}},
  \bibinfo {author} {\bibfnamefont {M.}~\bibnamefont {Ferrero}}, \bibinfo
  {author} {\bibfnamefont {E.~M.}\ \bibnamefont {Stoudenmire}}, \bibinfo
  {author} {\bibfnamefont {T.}~\bibnamefont {Sch\"afer}},\ and\ \bibinfo
  {author} {\bibfnamefont {A.}~\bibnamefont {Georges}},\ }\bibfield  {title}
  {\bibinfo {title} {Mott insulating states with competing orders in the
  triangular lattice hubbard model},\ }\href
  {https://doi.org/10.1103/PhysRevX.11.041013} {\bibfield  {journal} {\bibinfo
  {journal} {Phys. Rev. X}\ }\textbf {\bibinfo {volume} {11}},\ \bibinfo
  {pages} {041013} (\bibinfo {year} {2021})}\BibitemShut {NoStop}%
\bibitem [{\citenamefont {Perepelitsky}\ \emph {et~al.}(2016)\citenamefont
  {Perepelitsky}, \citenamefont {Galatas}, \citenamefont {Mravlje},
  \citenamefont {\ifmmode~\check{Z}\else \v{Z}\fi{}itko}, \citenamefont
  {Khatami}, \citenamefont {Shastry},\ and\ \citenamefont
  {Georges}}]{Peripelitsky16}%
  \BibitemOpen
  \bibfield  {author} {\bibinfo {author} {\bibfnamefont {E.}~\bibnamefont
  {Perepelitsky}}, \bibinfo {author} {\bibfnamefont {A.}~\bibnamefont
  {Galatas}}, \bibinfo {author} {\bibfnamefont {J.}~\bibnamefont {Mravlje}},
  \bibinfo {author} {\bibfnamefont {R.}~\bibnamefont {\ifmmode~\check{Z}\else
  \v{Z}\fi{}itko}}, \bibinfo {author} {\bibfnamefont {E.}~\bibnamefont
  {Khatami}}, \bibinfo {author} {\bibfnamefont {B.~S.}\ \bibnamefont
  {Shastry}},\ and\ \bibinfo {author} {\bibfnamefont {A.}~\bibnamefont
  {Georges}},\ }\bibfield  {title} {\bibinfo {title} {Transport and optical
  conductivity in the hubbard model: A high-temperature expansion
  perspective},\ }\href {https://doi.org/10.1103/PhysRevB.94.235115} {\bibfield
   {journal} {\bibinfo  {journal} {Phys. Rev. B}\ }\textbf {\bibinfo {volume}
  {94}},\ \bibinfo {pages} {235115} (\bibinfo {year} {2016})}\BibitemShut
  {NoStop}%
\bibitem [{\citenamefont {Vrani\ifmmode~\acute{c}\else \'{c}\fi{}}\ \emph
  {et~al.}(2020)\citenamefont {Vrani\ifmmode~\acute{c}\else \'{c}\fi{}},
  \citenamefont {Vu\ifmmode \check{c}\else \v{c}\fi{}i\ifmmode \check{c}\else
  \v{c}\fi{}evi\ifmmode~\acute{c}\else \'{c}\fi{}}, \citenamefont {Kokalj},
  \citenamefont {Skolimowski}, \citenamefont {\ifmmode~\check{Z}\else
  \v{Z}\fi{}itko}, \citenamefont {Mravlje},\ and\ \citenamefont
  {Tanaskovi\ifmmode~\acute{c}\else \'{c}\fi{}}}]{Vranic20}%
  \BibitemOpen
  \bibfield  {author} {\bibinfo {author} {\bibfnamefont {A.}~\bibnamefont
  {Vrani\ifmmode~\acute{c}\else \'{c}\fi{}}}, \bibinfo {author} {\bibfnamefont
  {J.}~\bibnamefont {Vu\ifmmode \check{c}\else \v{c}\fi{}i\ifmmode
  \check{c}\else \v{c}\fi{}evi\ifmmode~\acute{c}\else \'{c}\fi{}}}, \bibinfo
  {author} {\bibfnamefont {J.}~\bibnamefont {Kokalj}}, \bibinfo {author}
  {\bibfnamefont {J.}~\bibnamefont {Skolimowski}}, \bibinfo {author}
  {\bibfnamefont {R.}~\bibnamefont {\ifmmode~\check{Z}\else \v{Z}\fi{}itko}},
  \bibinfo {author} {\bibfnamefont {J.}~\bibnamefont {Mravlje}},\ and\ \bibinfo
  {author} {\bibfnamefont {D.}~\bibnamefont {Tanaskovi\ifmmode~\acute{c}\else
  \'{c}\fi{}}},\ }\bibfield  {title} {\bibinfo {title} {Charge transport in the
  hubbard model at high temperatures: Triangular versus square lattice},\
  }\href {https://doi.org/10.1103/PhysRevB.102.115142} {\bibfield  {journal}
  {\bibinfo  {journal} {Phys. Rev. B}\ }\textbf {\bibinfo {volume} {102}},\
  \bibinfo {pages} {115142} (\bibinfo {year} {2020})}\BibitemShut {NoStop}%
\bibitem [{Pin()}]{Pines66}%
  \BibitemOpen
  \href@noop {} {}\bibinfo {note} {See, e.g., D. Pines and P. Nozieres,
  {\textit{The Theory of Quantum Liquids}} (Avalon Publishing, 1999), Chapter
  1.5 and 4.7}\BibitemShut {NoStop}%
\bibitem [{Neg()}]{Negele98}%
  \BibitemOpen
  \href@noop {} {}\bibinfo {note} {See, e.g., John W. Negele and Henri Orland,
  {\textit{Quantum Many-Particle Systems}} (CRC Press, 1998), Chapter
  6.2}\BibitemShut {NoStop}%
\bibitem [{\citenamefont {Sharma}\ \emph {et~al.}(2021)\citenamefont {Sharma},
  \citenamefont {Principi},\ and\ \citenamefont {Maslov}}]{sharma2021optical}%
  \BibitemOpen
  \bibfield  {author} {\bibinfo {author} {\bibfnamefont {P.}~\bibnamefont
  {Sharma}}, \bibinfo {author} {\bibfnamefont {A.}~\bibnamefont {Principi}},\
  and\ \bibinfo {author} {\bibfnamefont {D.~L.}\ \bibnamefont {Maslov}},\
  }\bibfield  {title} {\bibinfo {title} {Optical conductivity of a dirac-fermi
  liquid},\ }\href {https://doi.org/10.1103/PhysRevB.104.045142} {\bibfield
  {journal} {\bibinfo  {journal} {Physical Review B}\ }\textbf {\bibinfo
  {volume} {104}},\ \bibinfo {pages} {045142} (\bibinfo {year}
  {2021})}\BibitemShut {NoStop}%
\bibitem [{\citenamefont {Jakli{\v{c}}}\ and\ \citenamefont
  {Prelov{\v{s}}ek}(2000)}]{jaklivc2000finite}%
  \BibitemOpen
  \bibfield  {author} {\bibinfo {author} {\bibfnamefont {J.}~\bibnamefont
  {Jakli{\v{c}}}}\ and\ \bibinfo {author} {\bibfnamefont {P.}~\bibnamefont
  {Prelov{\v{s}}ek}},\ }\bibfield  {title} {\bibinfo {title}
  {Finite-temperature properties of doped antiferromagnets},\ }\href
  {https://doi.org/10.1080/000187300243381} {\bibfield  {journal} {\bibinfo
  {journal} {Advances in Physics}\ }\textbf {\bibinfo {volume} {49}},\ \bibinfo
  {pages} {1} (\bibinfo {year} {2000})}\BibitemShut {NoStop}%
\bibitem [{mah()}]{mahan}%
  \BibitemOpen
  \href@noop {} {}\bibinfo {note} {See, e.g., G. D. Mahan,
  {\textit{Many-Particle Physics}} (Plenum, New York, 1983), Section
  8.1.B}\BibitemShut {NoStop}%
\bibitem [{\citenamefont {Rosch}\ and\ \citenamefont
  {Howell}(2005)}]{rosch2005zero}%
  \BibitemOpen
  \bibfield  {author} {\bibinfo {author} {\bibfnamefont {A.}~\bibnamefont
  {Rosch}}\ and\ \bibinfo {author} {\bibfnamefont {P.}~\bibnamefont {Howell}},\
  }\bibfield  {title} {\bibinfo {title} {Zero-temperature optical conductivity
  of ultraclean fermi liquids and superconductors},\ }\href
  {https://doi.org/10.1103/PhysRevB.72.104510} {\bibfield  {journal} {\bibinfo
  {journal} {Physical Review B}\ }\textbf {\bibinfo {volume} {72}},\ \bibinfo
  {pages} {104510} (\bibinfo {year} {2005})}\BibitemShut {NoStop}%
\bibitem [{Abr()}]{Abrikosov63}%
  \BibitemOpen
  \href@noop {} {}\bibinfo {note} {See, e.g., A. A. Abrikosov, L. Gorkov and I.
  E. Dzyaloshinski, {\textit{Methods Of Quantum Field Theory In Statistical
  Physics}} (Dover, 1963), Section 19.2}\BibitemShut {NoStop}%
\bibitem [{\citenamefont {Caprara}\ \emph {et~al.}(2007)\citenamefont
  {Caprara}, \citenamefont {Grilli}, \citenamefont {Di~Castro},\ and\
  \citenamefont {Enss}}]{Caprara07}%
  \BibitemOpen
  \bibfield  {author} {\bibinfo {author} {\bibfnamefont {S.}~\bibnamefont
  {Caprara}}, \bibinfo {author} {\bibfnamefont {M.}~\bibnamefont {Grilli}},
  \bibinfo {author} {\bibfnamefont {C.}~\bibnamefont {Di~Castro}},\ and\
  \bibinfo {author} {\bibfnamefont {T.}~\bibnamefont {Enss}},\ }\bibfield
  {title} {\bibinfo {title} {Optical conductivity near finite-wavelength
  quantum criticality},\ }\href {https://doi.org/10.1103/PhysRevB.75.140505}
  {\bibfield  {journal} {\bibinfo  {journal} {Phys. Rev. B}\ }\textbf {\bibinfo
  {volume} {75}},\ \bibinfo {pages} {140505} (\bibinfo {year}
  {2007})}\BibitemShut {NoStop}%
\end{thebibliography}%

\end{document}